# Gap Size Effects for the Kelvin-Helmholtz Instability in a Hele-Shaw Cell

L. Meignin[a], P. Ern[b], P. Gondret[a] and M. Rabaud[a]

[a] *Laboratoire Fluides, Automatique et Systèmes Thermiques, Universités P. & M. Curie et Paris-Sud, CNRS (UMR 7608), Bât 502, Campus Universitaire, 91405 Orsay Cedex, France.*

[b] *Institut de Mécanique des Fluides de Toulouse, CNRS (UMR 5502), Allée du Professeur Camille Soula, 31400 Toulouse, France.*

PACS numbers: 47.20.Ft, 47.20.Ma, 47.60.+i, 47.35.+i

We report experimental results for the Kelvin-Helmholtz instability between two immiscible fluids in parallel flow in a Hele-Shaw cell. We focus our interest on the influence of the gap size between the walls on the instability characteristics. Experimental results show that the instability threshold, the critical wavelength, the phase velocity and the spatial growth rate depend on this gap size. These results are compared to both the previous two-dimensional analysis of Gondret and Rabaud (1997) and the three-dimensional analysis recently derived by Plouraboué and Hinch (2001), showing that the agreement is still not complete especially when gap size increases.

# I. INTRODUCTION

The Kelvin-Helmholtz instability is a shear instability at the interface between two streams of fluids moving at different parallel velocities. The interface is destabilized by fluid inertia effect in competition with gravity and surface tension if any. This mechanism is presently investigated in the particular geometry of a Hele-Shaw cell which consists of two glass plates separated by a thin gap. In our set-up, a less dense and less viscous fluid (gas) is flowing horizontally and parallel above a more dense and more viscous fluid (viscous oil), the gap of the cell being transverse to both gravity and flow direction. Due to the thin gap, viscous dissipation at the walls is important and the basic flow obeys the Darcy law. The Hele-Shaw configuration insures that the shear is uniform in the streamwise direction and time independent [3], contrary to classical mixing layer set-ups. In a prior paper, the shear instability that arises at the interface between this co-current flow of gas and viscous oil has been presented both experimentally and analytically [1]. The theoretical analysis was two-dimensional (2D): the flow is assumed to remain Poiseuille like everywhere in the thinnest dimension and is then averaged through the gap of the cell. The instability threshold is shown to be equal to the prediction given by the classical inviscid analysis (Kelvin-Helmholtz instability) whereas the phase velocity is found to be in agreement with the value that would be yielded by a pure viscous analysis deriving from the Darcy law. In such an open flow, the transition from a convective to an absolute regime of the instability was also predicted analytically and characterized experimentally [4].

The 2D analysis has been recently improved by Ruyer-Quil [5] who derived the appropriate inertial corrections to the Darcy law by using a perturbative method and a polynomial expansion of the velocity field. This method was used successfully for films flowing down inclined planes [6]. Although the 2D analysis gives good magnitude orders (to be shown in Figs. 3, 5 and 6) and explains most of the phenomena observed, it cannot account for all the influences of the gap size noticed in the experiment. In this paper, experimental results concerning the influence of the gap size on the instability characteristics are compared to three-dimensional (3D) theoretical developments recently derived by Plouraboué and Hinch [2].

The rest of the paper is organized as follows. In Sec. II we summarize the 2D analysis of Ref. [1] recently improved by Ruyer-Quil [5] and also the new 3D analysis of Plouraboué and Hinch [2]. After briefly recalling the experimental set-up, we present in Sec. III the experimental measurements when varying the gap, i. e. the influence of the gap size on the onset of the instability, the phase velocity, the critical wavelength, and the spatial growth rate. Finally we compare them to theoretical values from both 2D and 3D analysis.

## II. THEORETICAL FRAMEWORK

### A. Two-dimensional analysis

In a Hele-Shaw cell (Fig. 1), the gap size $b$ is supposed to be small compared to the scale of the streamwise variation of the velocity field. Thus the transverse lubrication profile can be considered to evolve slowly in time and space. With an appropriate average of the Navier-Stokes equation through the gap, Ruyer-Quil [5] found that the equation for the gap averaged velocity $\bar{\mathbf{u}}$ is the following at first order in Reynolds number $\mathrm{Re} = \bar{U}b^2k\rho/\mu$:

$$\frac{6}{5}\frac{\partial \bar{\mathbf{u}}}{\partial t} + \frac{54}{35}\bar{\mathbf{u}}\cdot\nabla\bar{\mathbf{u}} = -\frac{1}{\rho}\nabla P - \mathbf{g} - \frac{12\mu}{\rho b^2}\bar{\mathbf{u}}, \tag{1}$$

where $\rho$ and $\mu$ are the density and the dynamic viscosity of the fluid, $\mathbf{g}$ is the gravity and $P$ is the pressure. In this equation, the viscous dissipative term $(12\mu/\rho b^2)\bar{\mathbf{u}}$ corresponds to a Darcy term. Equation (1) keeps the same form than the relation derived in Ref. [1] but with more exact numerical coefficients for the inertial and unstationary terms. The dispersion relation for perturbations $\exp[i(kx - \omega t)]$ that can be derived is thus the following:

$$\frac{6}{5}\omega^2 - \left[\frac{96}{35}\frac{(\rho_1\bar{U}_1 + \rho_2\bar{U}_2)k}{\rho_1 + \rho_2} - i\frac{12(\mu_1 + \mu_2)}{(\rho_1 + \rho_2)b^2}\right]\omega + \frac{54}{35}\frac{(\rho_1\bar{U}_1^2 + \rho_2\bar{U}_2^2)k^2}{\rho_1 + \rho_2} - \frac{(\rho_2 - \rho_1)gk + \gamma\frac{\pi}{4}k^3}{\rho_1 + \rho_2}$$

$$- i\frac{12(\mu_1\bar{U}_1 + \mu_2\bar{U}_2)k}{(\rho_1 + \rho_2)b^2} = 0, \tag{2}$$

where $\bar{U}_i$ is the mean basic velocity in each fluid, $\gamma$ is the interfacial tension, $k$ the wavenumber and $\omega$ the frequency, and subscripts 1 and 2 stand for gas and oil.

This dispersion relation provides the onset of instability for the gas velocity, the phase velocity and the wavelength at onset, and also the temporal growth rate, which are respectively given by the following expressions in the zero viscosity ratio limit ($\mu_1/\mu_2 \to 0$) corresponding to the experiment:

$$\overline{U}_{1c}^{2D} = \frac{35}{54}\left\{\frac{[\pi(\rho_2-\rho_1)g\gamma]^{1/2}}{\rho_1}\right\}^{1/2},$$

$$V_\phi^{2D} = \frac{2\mu_1 \overline{U}_{1c}^{2D}}{\mu_2},$$

$$\lambda_c^{2D} = 2\pi\left[\frac{\pi\gamma}{4(\rho_2-\rho_1)g}\right]^{1/2},$$

$$\omega_i^{2D} = \frac{b^2 k^2}{10\mu_2}\left[\frac{9}{7}\rho_1 \overline{U}_1^2 - \frac{5}{6}\frac{(\rho_2-\rho_1)g + \frac{\pi\gamma k^2}{4}}{\rho_1}\right].$$

Note that the gap size $b$ does not appear anymore in these expressions, except for the temporal growth rate.

### B. Three-dimensional analysis

A 3D analysis has been recently developed by Plouraboué and Hinch [2], where unstationary terms are allowed to deviate from the Poiseuille profile. Starting from the full Navier-Stokes equation, the gap averaged description is replaced by an asymptotic analysis by considering the gap size $b$ small compared to the critical wavelength $\lambda_c$ ($b/\lambda_c \ll 1$). The basic flow, both in gas and oil, is still supposed to be parabolic and a stationary perturbed interface is considered. In the small viscosity ratio limit ($\mu_1/\mu_2 \ll 1$), the velocity perturbation remains also parabolic in oil as inertia can be neglected in this viscous fluid. However inertia terms are taken into account in the gas and the perturbation for the gas velocity deviates from the parabolic profile. Using the lubrication hypothesis and linearizing the Navier-Stokes equation, the pressure is found to be potential for the two fluids. Thanks to the boundary conditions at the interface, i.e. the usual dynamic condition for the pressure and the kinematic condition written for the gap averaged velocity, Plouraboué and Hinch [2] derived a differential equation which links the velocity profile and the pressure. Their equation is dimensionless and the gap size $b$ does not appear any more except in the gas Reynolds number $\text{Re}_1 = \overline{U}_1 b^2 k \rho_1/\mu_1$. In order to compare with our measurements, we write a dimensional version of their differential equation for the velocity perturbation in the gas $u_1$:

$$-i\omega \frac{u_1}{\bar{u}_1} + ik\bar{U}_1 \frac{3}{2}\left[1 - \left(\frac{2z}{b}\right)^2\right] \frac{u_1}{\bar{u}_1} - \frac{\mu_1}{\rho_1}\frac{\partial^2}{\partial z^2}\frac{u_1}{\bar{u}_1}$$
$$= \frac{k}{\rho_1}\left[(\rho_1 - \rho_2)g - \frac{\pi}{4}\gamma k^2 + \frac{12i\omega\mu_2}{kb^2} - \frac{12i\mu_1\bar{U}_1}{b^2}\right]\frac{1}{-i\omega + ik\bar{U}_1} \quad (3),$$

where $\bar{U}_1$ is the mean gas velocity, $z$ the transverse direction and $\bar{u}_1 = \frac{1}{b}\int_{-b/2}^{b/2} u_1(z)dz$ corresponds to the average of $u_1$ over the gap, together with the boundary conditions at the walls $u_1(\pm b/2) = 0$. Note that in the limit $b \to 0$, one recovers the 2D analysis.

We solve Eq. (3) with a standard shooting method and find a solution $u_1(z)$ which depends on $k$ and $\omega$ (all the other parameters are known except the mean gas velocity $\bar{U}_1$ which is tuneable numerically). In the temporal approach, $k$ is taken real and $\omega$ complex: $\omega = \omega_r + i\omega_i$. Obviously not all the couples $(k,\omega)$ are solutions of the problem and the last condition we have to verify is the normalization of $u_1(z)$: $\bar{u}_1 = 1$. When setting $\bar{U}_1$ to a low value, all the couples $(k,\omega)$ verify $\omega_i < 0$ and the system is then stable. By increasing $\bar{U}_1$, a band of couples verifies $\omega_i > 0$ and the system is thus unstable. At the onset of the instability $\bar{U}_{1c}$, only one couple $(k,\omega) = (k_c,\omega_r)$ satisfies $\omega_i = 0$. The phase velocity $V_\phi = \omega_r / k_c$ and the wavelength $\lambda_c = 2\pi / k_c$, both at threshold, are also determined, together with an estimation of the spatial growth rate $k_i = \omega_i / V_\phi$ just above the threshold.

We solve numerically this problem for values of the gap size $b$ varying from 0.125 mm to 1 mm (problems of convergence for the algorithm were encountered outside this range). The results of these calculations will be presented in Figs. 3, 5 and 6, and compared to the experimental values.

The velocity perturbation profile $u_1(z)$ at the onset $\bar{U}_{1c} = 3.58$ m/s for $b = 0.350$ mm is presented for two different frequencies in Fig. 2. This profile is very close to a parabola for $\omega_r = 2.5$ rad/s and becomes more flat for $\omega_r = 5.1$ rad/s, the most unstable frequency at this gap size. The skin effect in the gas (Stokes layers) on the typical scale $\delta = \sqrt{\mu_1 / \rho_1 \omega_r}$ [7] is here quite weak ($\delta / b = 4.7$ for $\omega_r = 5.1$ rad/s). For larger $\bar{U}_1$ and $\omega_r$, the velocity perturbation profile could deviate strongly from the parabolic profile with a slower middle flow and the appearance of two separate boundary layers.

## III. EXPERIMENTS AND DISCUSSION

### A. Experimental set-up

The cell is made of two glass plates separated by a thin gap $b$ which is varied between 0.175 mm and 0.830 mm. The gas (nitrogen of viscosity $\mu_1 = 1.75 \; 10^{-5}$ Pa.s and density $\rho_1 = 1.28$ kg/m³ at the temperature T = 20°C and pressure $P = 1.1 \; 10^5$ Pa) and the liquid (silicon oil of viscosity $\mu_2 = 0.02$ Pa.s, density $\rho_2 = 952$ kg/m³ and interfacial tension $\gamma = 0.02$ N/m at the temperature T = 20°C) flow through the cell as shown in Fig. 1. The experimental set-up has been presented in detail in Ref. [1]. Gravity acts in the plane of the cell, perpendicular to the interface. Both fluids are injected at the same controlled pressure $P_{in}$ and get out at atmospheric pressure $P_{atm}$. The pressure difference $\Delta P = P_{in} - P_{atm}$ is our experimental control parameter. Below the instability threshold (low $\Delta P$), the two fluids flow parallel with horizontal interface and obey the Darcy law $\overline{U}_i = \left(b^2 / 12 \mu_i\right)(\Delta P / L)$ where $L$ is the length of the cell. The pressure gradient $\Delta P / L$ being the same for the two fluids, their velocities $\overline{U}_1$ and $\overline{U}_2$ are linked by the relation $\mu_1 \overline{U}_1 = \mu_2 \overline{U}_2$. Because of the strong viscosity contrast, oil is thus flowing slowly compared to gas ($\overline{U}_2 \approx 5$ mm/s $<<$ $\overline{U}_1 \approx 5$ m/s). From the accurate measurement of the pressure difference $\Delta P$, we deduce $\overline{U}_1$ from the Darcy law.

By applying a small periodic modulation of the oil injection pressure, with an amplitude and a frequency controlled by an electrovalve, the interface develops periodic deformation locally at the end of the splitter tongue. The perturbation thus created is either damped (below threshold), marginal (at threshold) or amplified (above threshold), depending on the values of both forcing frequency $f$ and gas velocity $\overline{U}_1$ (see Fig. 2 of Ref. [4]). This behavior allows us to build the experimental marginal stability curve (see Fig. 4 of Ref. [1]). The onset of the instability for the gas velocity $\overline{U}_{1c}$ is defined as the minimum of this curve. All the onset determinations have been made at small forcing amplitude (less than 0.5 mm). For example, at $b = 0.350$ mm, the onset corresponds to $\overline{U}_{1c} = 4.22$ m/s with the most unstable frequency $f_c = 0.4$ Hz and the critical wavelength $\lambda_c \approx 1.15$ cm. For a different frequency (lower or upper than $f_c$) the unstable state is obtained for larger $\overline{U}_1$.

### B. Influence of the gap size

In the present work, measurements for six different gap values have been performed. Experimentally, the gap range is limited at large $b$ ($b = 0.830$ mm) and at small $b$ ($b = 0.175$ mm) by the range of our pressure regulation. For all these thicknesses, the instability threshold $\overline{U}_{1c}$, the phase velocity $V_\phi$ and the wavelength $\lambda_c$ at onset and also the spatial growth rate $k_i$ just above threshold are measured.

### *1. Instability threshold*

The evolution of the experimental onset $\overline{U}_{1c}$ with the gap is presented in Fig. 3 as well as the results derived from the 2D and 3D theories. The experimental onset $\overline{U}_{1c}$ increases slightly with $b$ up to $b = 0.4$ mm and strongly for larger $b$. The predicted value deriving from the 2D analysis is independent of the gap size ($\overline{U}_{1c}^{2D} = 3.50$ m/s) and slightly smaller than the experimental values. The 3D analysis also shows an increase for the instability threshold with the gap size, though the whole increase is quite smaller than the experimental one. The onset is systematically underestimated of about 15% for small gap and 50% for larger gap. This disagreement could perhaps be ascribed either to the effect of the boundary layer existing near the interface between gas and oil or to the fact that theories do not take into full account the semi-circular shape of the meniscus in the transverse direction. Indeed these effects could become important if the condition $b/\lambda_c << 1$ is not well satisfied. Note that at low $b$, the injection pressure is large and compressibility effect thus becomes more important. If taken into account, this effect will decrease the experimental evolution of $\overline{U}_{1c}$: by 17% at $b = 0.175$ mm and by 1% at $b = 0.830$ mm. Thus at the smallest gap the agreement between theory and experiment is much better but disagreement still exists for the largest gap.

### *2. Phase velocity*

The phase velocity $V_\phi$ of the waves is also determined for the different gaps at the corresponding instability threshold $\overline{U}_{1c}$. We restrict our study to the linear regime, i.e. approximately the first centimeters of the interface near inlet when wave amplitude is smaller than 1.5 mm. Spatio-temporal graphs $(x,t)$ have been built by recording a video line parallel to the interface as a function of time and $V_\phi$ is calculated as the local slope of the wave trajectories (Fig. 4). Spatial averaged values of $V_\phi$ on the first centimeters are displayed in Fig. 5a and the reported error bars correspond to the extrema measured values. The phase velocity increases with the gap thickness from typically 5 mm/s at $b = 0.175$ mm to 18 mm/s at $b = 0.830$ mm. As $\overline{U}_{1c}$ also increases with the gap $b$, the ratio $V_\phi / \overline{U}_{1c}$ as a function of $b$ is plotted in Fig. 5b. As this ratio still increases with $b$, the evolution of the phase velocity with the gap is not directly related to the increase of the onset, but can be considered as intrinsic. The increase of $V_\phi$ at larger gaps may be due to relatively less important friction of the waves on the walls. For the 3D theory, the evolution for both $V_\phi$ and $V_\phi / \overline{U}_{1c}$ is very similar to the evolution of the experimental values but with only slightly higher values (Fig. 5). A possible improvement would be to consider the dissipation in the liquid wall films associated with the wave motion [2]. Concerning the 2D theory, the predictions of a constant phase velocity $V_\phi^{2D} = 6.125$ mm/s and a constant ratio $V_\phi^{2D} / \overline{U}_{1c}^{2D} = 1.7510^{-3}$ are rather good for small $b$ but fail at larger $b$ (Fig. 5).

### *3. Wavelength*

Thanks to recorded pictures of the interface, the wavelength $\lambda_c$ at onset for the most unstable frequency $f_c$ has also been measured directly for the different gap sizes and is plotted in Fig. 6a. The experimental wavelength increases with the gap, in agreement with the observed increase in $V_\phi$ and the slight decrease of $f_c$ from 0.43 Hz to 0.35 Hz ($\lambda_c = V_\phi / f_c$). The value derived from the 2D analysis $\lambda_c^{2D} = 0.815$ cm is slightly lower than the experimental ones at small gap but yields a good magnitude order. The values calculated from the 3D analysis increase with the gap as in the experiment but much more weakly.

### *4. Spatial growth rate*

Finally we have determined for each gap size the spatial growth rate $k_i$ just above threshold at $\overline{U}_1 = 1.03 \, \overline{U}_{1c}$. For that, we fit the shape of the wave during its spatial amplification by the product of an exponentially growing amplitude, $\exp(k_i x)$, by an oscillatory signal of constant amplitude (Fig. 7). In Fig. 6b, the experimental spatial growth rate $k_i$ appears to be nearly constant at the value 18 m$^{-1}$. The uncertainty for this value is important (15%) because even so close to threshold, there is only a few arches of sinusoid to determine $k_i$ before non-linear saturation and important change of $\lambda$. For the 2D analysis, we consider $k_i^{2D} = \omega_i^{2D} / V_\phi^{2D}$ with $k_r = k_c$ for the calculation of $\omega_i^{2D}$ 3% above $\overline{U}_{1c}^{2D}$ and the set of points is thus on a parabola. Although the values are close to the experimental ones at low $b$, the 2D analysis strongly overestimates $k_i$ for larger $b$ (Fig. 6b). The theoretical estimation of the spatial growth rate obtained with the 3D analysis 3% above theoretical onset is better but still increases with the gap too strongly compared to the experimental values (Fig. 6b); for the largest gap, the theoretical value is five times higher than the experimental one. The three-dimensional analysis fails to reproduce precisely the evolution of both wavelength and spatial growth rate even if it improves the results of the 2D analysis. We have no physical argument to explain why the agreement is so bad on these variables.

## 5. Validity of the approximations

Taking the reduced gas Reynolds number $\text{Re}_1 = \overline{U}_1 b^2 k \rho_1 / \mu_1$ containing the two length scales $b$ and $1/k$ [2] there is a good agreement between the experiments and the 3D theory for the threshold value $\text{Re}_{1c} = \overline{U}_{1c} b^2 k_c \rho_1 / \mu_1$ at small $b$, as for the 2D theory. However, as $b$ increases the theoretical value becomes higher than the experimental one: e.g., for $b = 0.830$ mm, $\text{Re}_{1c} = 131$ theoretically whereas $\text{Re}_{1c} = 88$ experimentally. This difference is possibly related to the values of the parameter $kb$ which are supposed to be small compared to 1 in the theories, although the 3D predictions are not completely self-consistent with this hypothesis. In Fig. 8, the product $k_c b$ (value of $kb$ at onset) as a function of $b$ is displayed: the condition $k_c b << 1$ is correctly satisfied at each gap size in the experiment but not at large $b$ for both 2D and 3D analysis. A second point is that theoretical values for the most unstable frequency are different from the one found experimentally: e.g., at the onset $\overline{U}_{1c} = 3.58$ m/s for $b = 0.350$ mm, $\omega_{r\,\text{exp}} = 2.5$ rad/s and $\omega_{r\,2D} = 2\pi V_\phi^{2D} / \lambda_c^{2D} = 4.7$ rad/s or $\omega_{r\,3D} = 5.1$ rad/s. With $\omega_r = 5.1$ rad/s, the 3D theory predicts a shape of the velocity perturbation different from a parabola whereas with $\omega_r = 2.5$ rad/s the profile is closer (Fig. 2). The 3D theory then overestimates the perturbation to the Poiseuille profile.

## 6. Non-linear shape of the interface

In the results presented above, we have only considered the linear domain. In our set-up the waves remain linear only along the first centimeters of the cell. The non-linear effects appear quite rapidly and the waves saturate further. The up/down symmetry is broken and bumps of the viscous phase propagate on an almost flat interface. The amplitude of the saturated waves is about 3 mm whatever the gap size is. Snapshots of saturated waves are presented for two different gaps in Fig. 9: whereas the amplitude remains roughly the same, the shape is quite different. However theoretical non-linear predictions on this aspect still remain to be done.

## IV. CONCLUSION

In an Hele-Shaw set-up, we study in detail the mechanisms of the Kelvin-Helmholtz instability when transverse dissipative terms are important. The experimental results show that the instability characteristics —namely the instability threshold, the critical wavelength, the phase velocity and the spatial growth rate— strongly increase with the gap size of the Hele-Shaw cell. In the case of the phase velocity, the evolution with the gap size is close to the three-dimensional analysis predictions. However, for the other measured quantities, the agreement between experiment and theories existing for the smallest gaps is only qualitative for the largest gaps. This difference may be due to the theoretical assumption $b/\lambda_c <<1$ which is no more well verified at large $b$ for both theories. The three-dimensional approach improves the two-dimensional analysis as it gives the good qualitative evolution of the instability characteristics with the gap size. The agreement is not quantitatively satisfactory and a more complete analysis appears therefore necessary, possibly including the shape of the meniscus between gas and oil and the dissipation in the liquid wall films as underlined in Ref. [2]. The next necessary step is a non-linear study of the instability, both experimentally and theoretically.

## ACKNOWLEDGEMENTS

The authors acknowledge C. Ruyer-Quil, P.-Y. Lagrée, M. Rossi, E. J. Hinch and F. Plouraboué for fruitful discussions.

# Figure Captions

FIG. 1. Sketch of the experimental set-up, the thickness $b$ varying from 0.175 mm to 0.830 mm.

FIG. 2. Profile of the velocity perturbation in the gas $u_1(z)$ through the gap ($b = 0.350$ mm and $\overline{U}_1 = \overline{U}_{1c} = 3.58$ m/s): $\omega_r = 2.5$ rad/s (○), $\omega_r = 5.1$ rad/s (+) and parabolic profile (——).

FIG. 3. Critical gas velocity at onset $\overline{U}_{1c}$ versus the gap size $b$: experimental values (●) with error bars, 3D theory (—⊖—), 2D theory (- -).

FIG. 4. Typical spatio-temporal graph showing the wave trajectories at $\overline{U}_{1c} = 4.3$ m/s and $b = 0.250$ mm (a black line represents the trajectory of a point of constant height of the wave). An interface length of about 0.5 cm is displayed during 5 s.

FIG. 5. a) Phase velocity $V_\phi$ of the waves at onset versus the gap size $b$, b) Ratio of the phase velocity $V_\phi$ to the gas velocity at onset $\overline{U}_{1c}$ versus the gap size $b$. Experimental values (●), 3D theory (—⊖—), 2D theory (- -).

FIG. 6. a) Evolution of the wavelength $\lambda_c$ at onset versus the gap thickness $b$, b) Evolution of the spatial growth rate $k_i$ at $\overline{U}_1 = 1.03\,\overline{U}_{1c}$ versus the gap thickness $b$. Experimental values (●), 3D theory (—⊖—), 2D theory (- -).

FIG. 7. Typical spatial evolution of the amplitude A of a wave in the linear regime at a given time for $b = 0.250$ mm and $\overline{U}_1 = 1.03\,\overline{U}_{1c}$: experimental points (○) and exponential fit (——) ($k_i = 18.5$ m$^{-1}$ and $k_r$ 650 m$^{-1}$). $x$ is the distance downstream from the beginning of the interface. Note that experimentally the wavelength increases slightly with $x$ and thus for the fit, the wavenumber $k_r$ is allowed to decrease slightly with $x$.

FIG. 8. Evolution of the product $k_c b$ versus the gap thickness $b$. Experimental values (●), 3D theory (—⊙—), 2D theory (- -).

FIG. 9. Snapshots of waves of saturated amplitude propagating to the right: a) $b = 0.350$ mm, b) $b = 0.830$ mm.

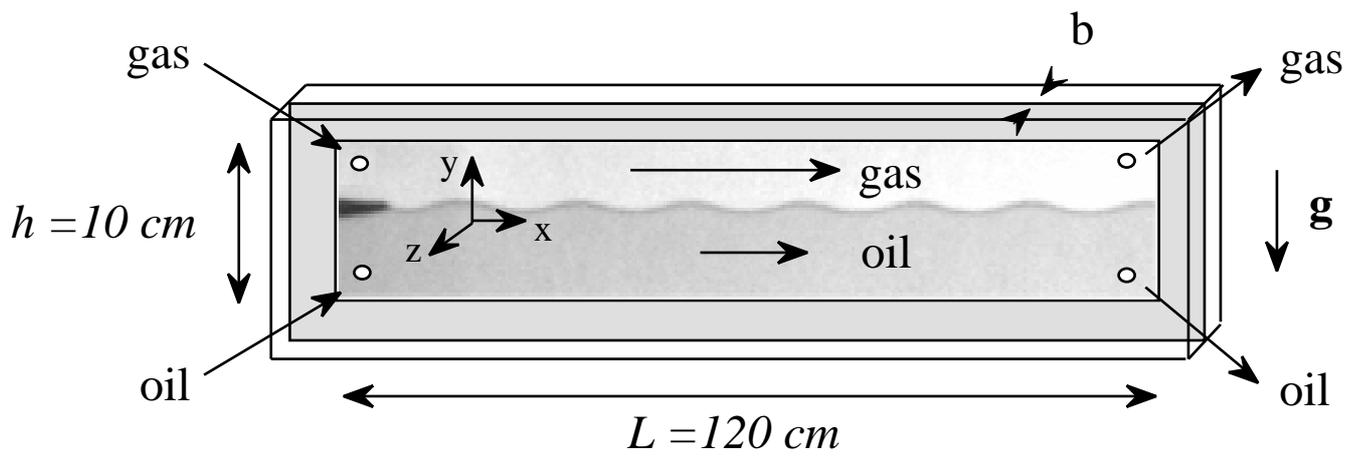

FIG. 1. Sketch of the experimental set-up, the thickness $b$ varying from 0.175 mm to 0.830 mm.

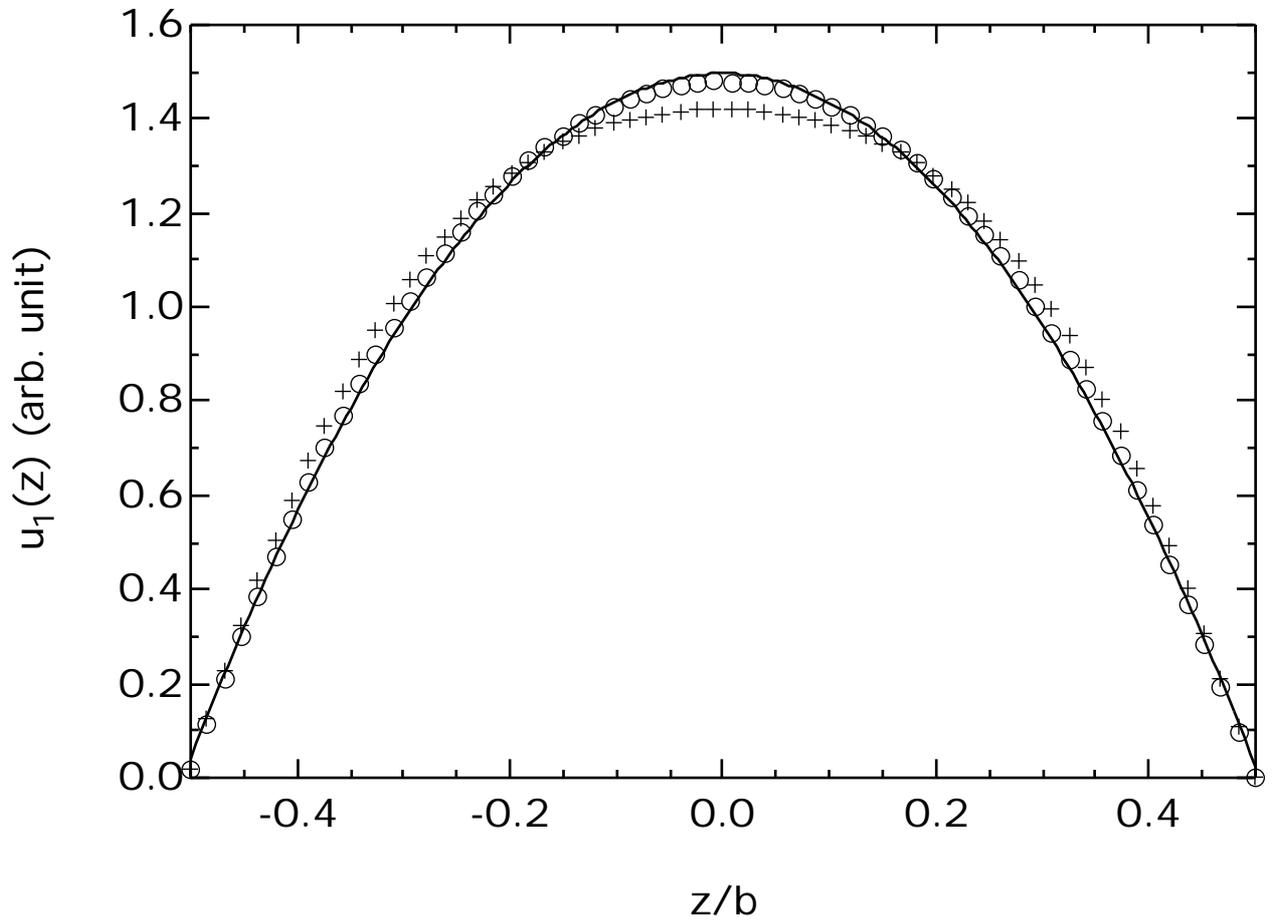

FIG. 2. Profile of the velocity perturbation in the gas $u_1(z)$ through the gap ($b = 0.350$ mm and $\overline{U}_1 = \overline{U}_{1c} = 3.58$ m/s): $\omega_r = 2.5$ rad/s (○), $\omega_r = 5.1$ rad/s (+) and parabolic profile (——).
MEIGNIN, Physical Review E

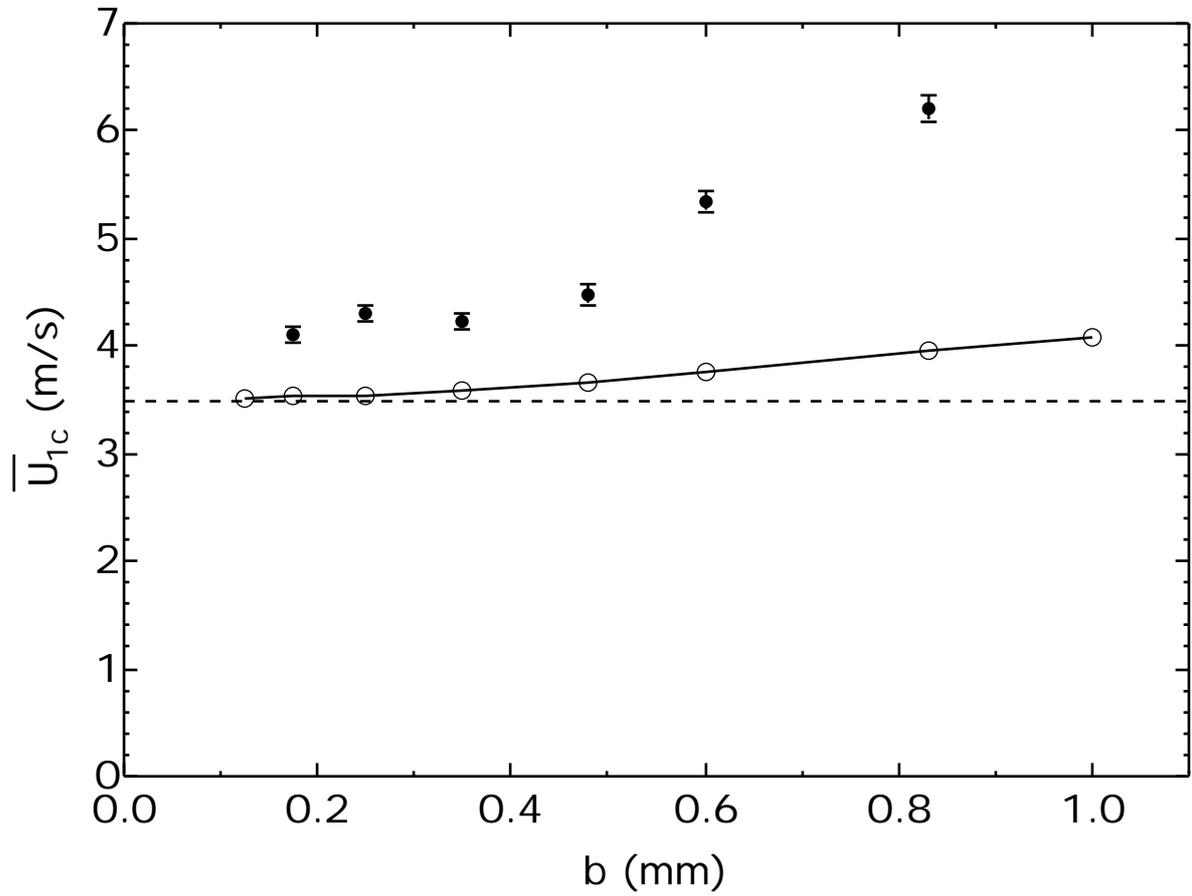

FIG. 3. Critical gas velocity at onset $\overline{U}_{1c}$ versus the gap size $b$: experimental values (●) with error bars, 3D theory (−○−), 2D theory (- -).

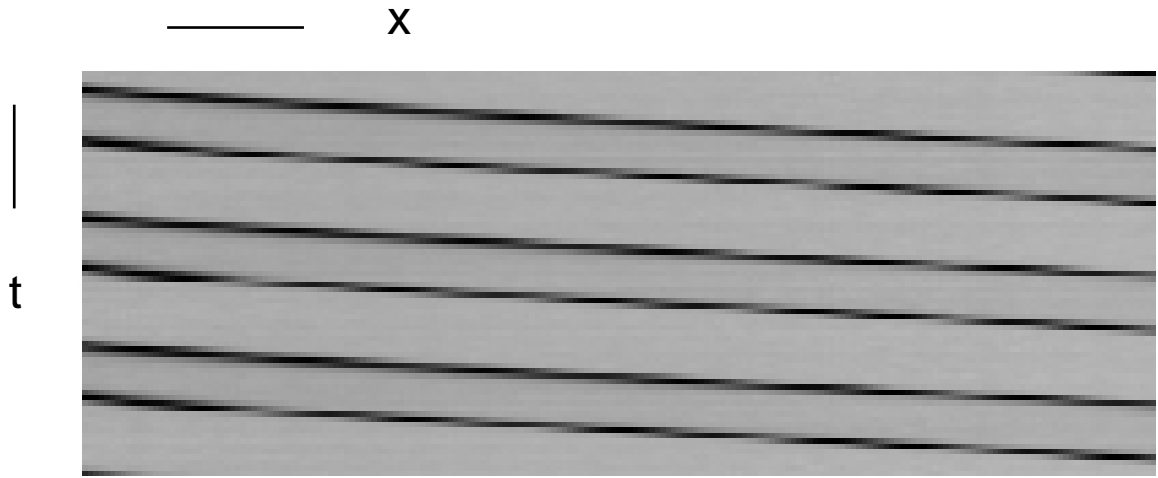

FIG. 4. Typical spatio-temporal graph showing the wave trajectories at $\overline{U}_{1c} = 4.3$ m/s and $b = 0.250$ mm (a black line represents the trajectory of a point of constant height of the wave). An interface length of about 0.5 cm is displayed during 5 s.

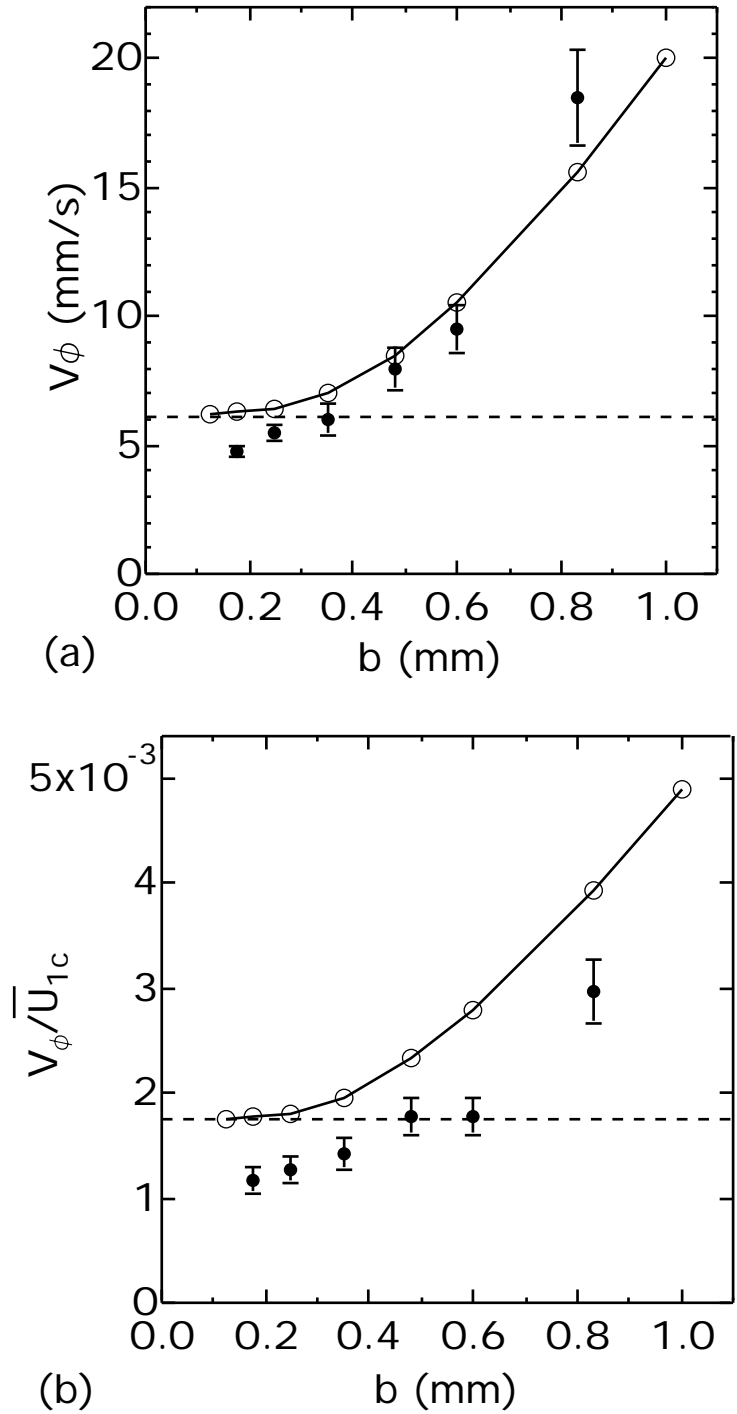

FIG. 5. a) Phase velocity $V_\phi$ of the waves at onset versus the gap size $b$, b) Ratio of the phase velocity $V_\phi$ to the gas velocity at onset $\overline{U}_{1c}$ versus the gap size $b$. Experimental values (●), 3D theory (–o–), 2D theory (- -).

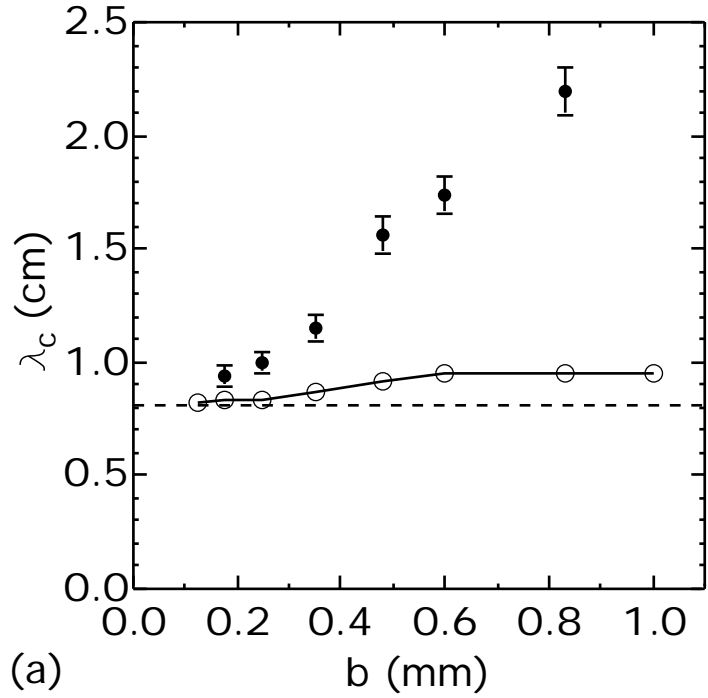

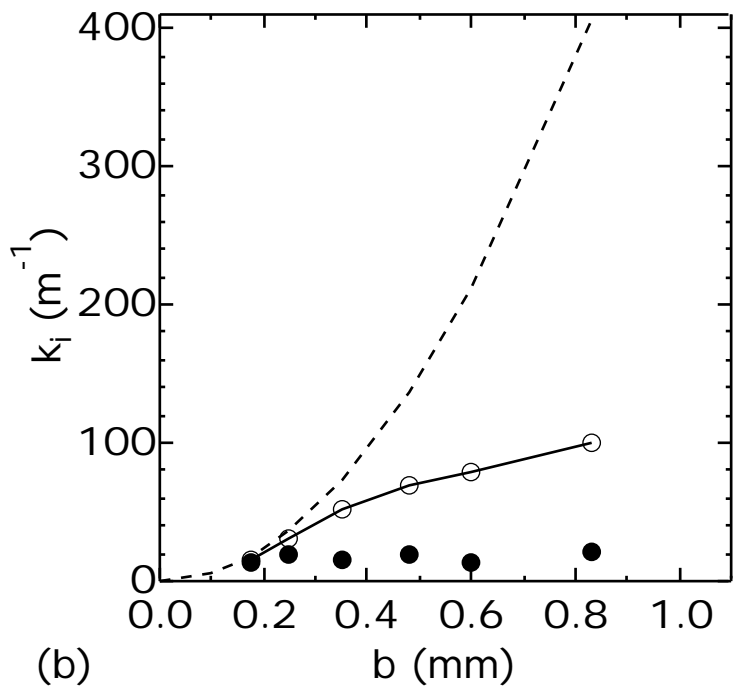

FIG. 6. a) Evolution of the wavelength $\lambda_c$ at onset versus the gap thickness $b$, b) Evolution of the spatial growth rate $k_i$ at $\overline{U}_1 = 1.03\,\overline{U}_{1c}$ versus the gap thickness $b$. Experimental values (●), 3D theory (─○─), 2D theory (- -).

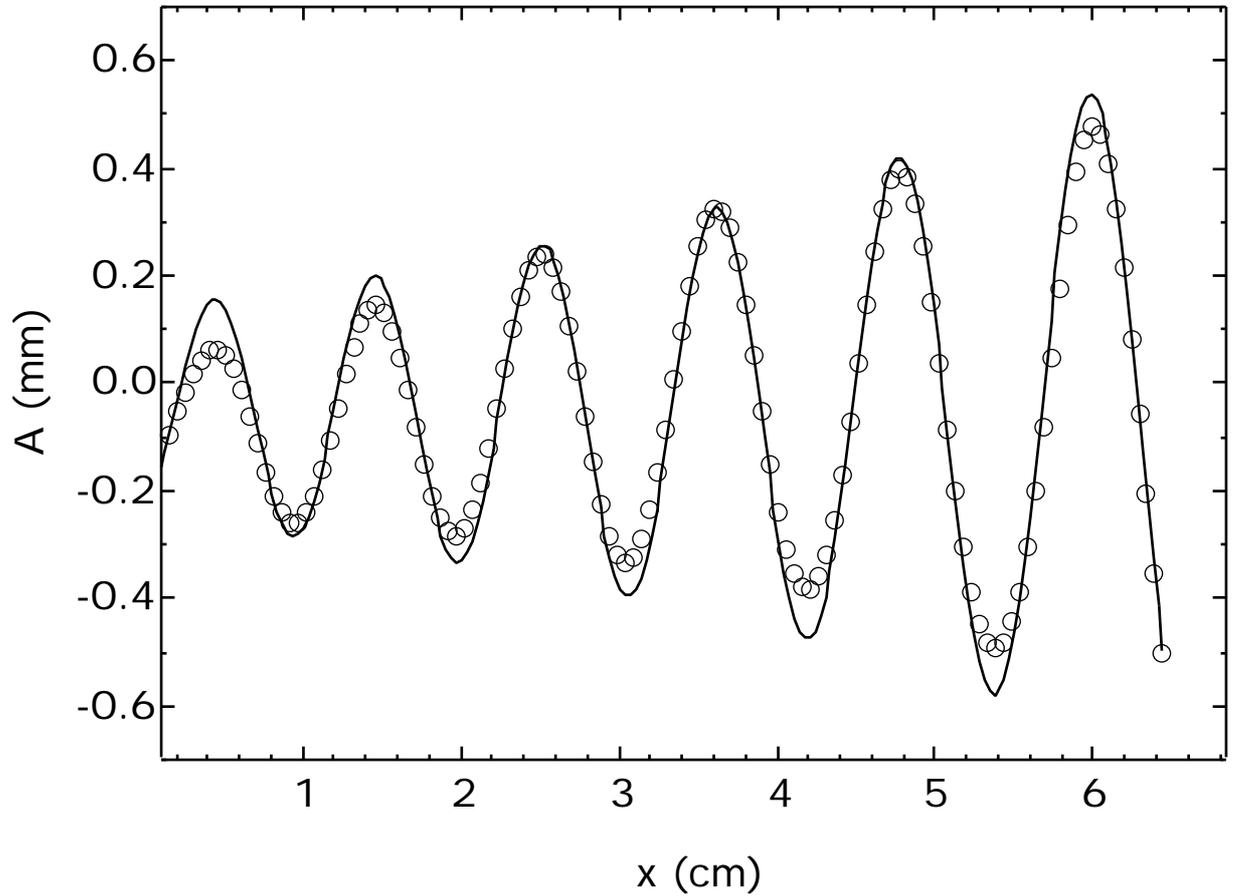

FIG. 7. Typical spatial evolution of the amplitude A of a wave in the linear regime at a given time for $b = 0.250$ mm and $\overline{U}_1 = 1.03\, \overline{U}_{1c}$: experimental points (○) and exponential fit (——) ($k_i = 18.5$ m$^{-1}$ and $k_r$ 650 m$^{-1}$). $x$ is the distance downstream from the beginning of the interface. Note that experimentally the wavelength increases slightly with $x$ and thus for the fit, the wavenumber $k_r$ is allowed to decrease slightly with $x$.

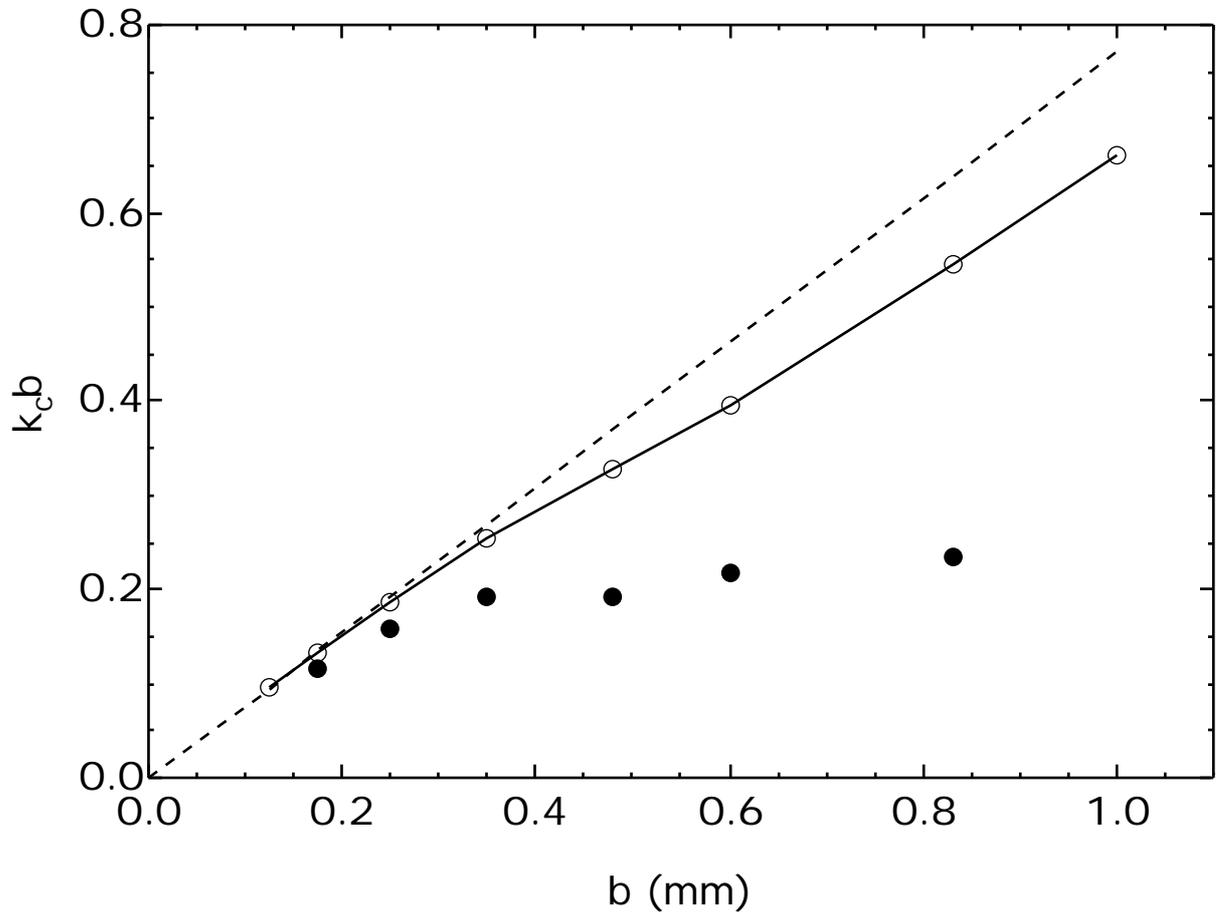

FIG. 8. Evolution of the product $k_c b$ versus the gap thickness $b$. Experimental values (●), 3D theory (–○–), 2D theory (- -).

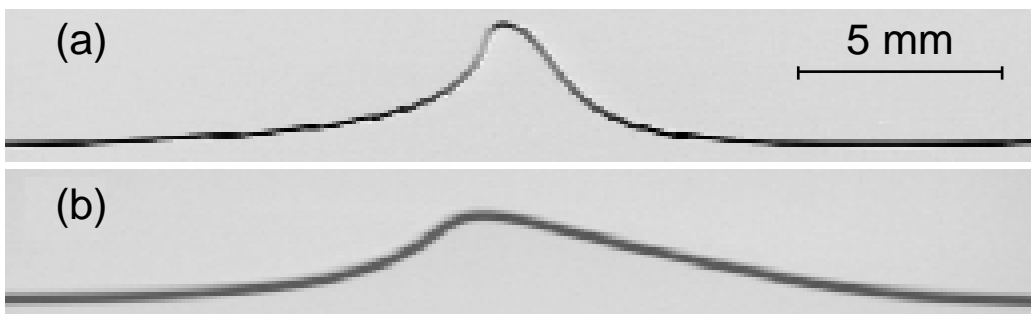

FIG. 9. Snapshots of waves of saturated amplitude propagating to the right: a) $b = 0.350$ mm, b) $b = 0.830$ mm.